\documentclass[conference]{IEEEtran}
\IEEEoverridecommandlockouts
\usepackage{amsmath,amssymb,amsfonts}
\usepackage{graphicx}
\usepackage{textcomp}
\usepackage{xcolor}

\usepackage{hyperref}
\usepackage{cleveref}

\usepackage{algorithm}
\usepackage{algorithmicx}
\usepackage{algpseudocode}
\usepackage{blkarray}
\usepackage{graphicx}
\graphicspath{ {./images/} }

\def\BibTeX{{\rm B\kern-.05em{\sc i\kern-.025em b}\kern-.08em
    T\kern-.1667em\lower.7ex\hbox{E}\kern-.125emX}}

\usepackage[style=ieee]{biblatex}
\usepackage{tikz}
\usetikzlibrary{shapes.geometric, arrows}
\usepackage{bera}
\usepackage{listings}
\usepackage{xcolor}

\colorlet{punct}{red!60!black}
\definecolor{background}{HTML}{EEEEEE}
\definecolor{delim}{RGB}{20,105,176}
\definecolor{numb}{RGB}{106, 109, 32}
\definecolor{string}{rgb}{0.64,0.08,0.08}

\lstdefinelanguage{json}{
    numbers=left,
    numberstyle=\small,
    rulecolor=\color{black},
    showspaces=false,
    showtabs=false,
    breaklines=true,
    firstnumber=1,
    xleftmargin=3.2ex,
    postbreak=\raisebox{0ex}[0ex][0ex]{\ensuremath{\color{gray}\hookrightarrow\space}},
    breakatwhitespace=true,
    basicstyle=\ttfamily\small,
    upquote=true,
    morestring=[b]",
    stringstyle=\color{string},
    literate=
     *{0}{{{\color{numb}0}}}{1}
      {1}{{{\color{numb}1}}}{1}
      {2}{{{\color{numb}2}}}{1}
      {3}{{{\color{numb}3}}}{1}
      {4}{{{\color{numb}4}}}{1}
      {5}{{{\color{numb}5}}}{1}
      {6}{{{\color{numb}6}}}{1}
      {7}{{{\color{numb}7}}}{1}
      {8}{{{\color{numb}8}}}{1}
      {9}{{{\color{numb}9}}}{1}
      {\{}{{{\color{delim}{\{}}}}{1}
      {\}}{{{\color{delim}{\}}}}}{1}
      {[}{{{\color{delim}{[}}}}{1}
      {]}{{{\color{delim}{]}}}}{1},
}

\begin{document}

\title{An adaptable JSON Diff Framework}

\author{
\IEEEauthorblockN{Ao Sun}
}

\maketitle

\begin{abstract}
In this paper, we present an implementation of JSON-diff framework JYCM, extending the existing framework by introducing the concept of "unordered" comparisons and allowing users to customize their comparison scenarios flexibly. Furthermore, we provide a diff-result renderer to visualize better and understand the differences between JSON objects. Our work enables more adaptable and comprehensive comparisons to accommodate a wider range of use cases and requirements.
\end{abstract}


\section{Introduction}
JSON \cite{crockford2006json} as a protocol has become prevalent in web applications \cite{pezoa2016foundations}, where it is used as the most common input and output format. Many test cases have been created based on that to ensure the quality assurance of large web applications. A typical use case is that we periodically send JSON input to an idempotent web API and obtain JSON output, then use scripts to verify whether the output JSON meets expectations by differing the output with a target one, which is usually called JSON diff. As these technologies are increasingly used, we face several challenges.

First, the returned JSON often contains fields like timestamp that need to be excluded naturally during the JSON diff process. We note that these fields are not always on the top level; they could appear anywhere in JSON. Second, in scenarios involving large JSON composed of complex nested structures and long-length arrays, it can be challenging for users to view and analyze JSON diff. Third, for a particular field in JSON, whether it "has been changed or its change is OK" really depends on the context. For example, for an API that outputs the bounding box of a person in a picture based on an input image URL, usually, we should compare it with the benchmark bounding box in IOU metrics rather than comparing the four coordinates of the bounding box are the same or not. Additionally, when an API returns an array, its meaning in the context could be a set, so it should be compared as two sets instead of ordered arrays.

In this paper, we have proposed a JSON diff framework \href{(https://github.com/eggachecat/jycm}{JYCM}, which you can access on \href{(https://github.com/eggachecat/jycm}{github} \footnote{https://github.com/eggachecat/jycm}, several contributions to the field of JSON diff:

\begin{itemize}
    \item We have presented an implementation of a JSON diff framework, demonstrating high efficiency, adaptability, and scalability for various use cases.
    
    \item Our framework introduces the concept of unordered comparison for JSON arrays, which enhances its applicability in scenarios where the order of elements is not crucial.
    
    \item Our framework allows users to flexibly customize the comparison logic according to their specific requirements, such as comparing or matching only the IDs of objects within a collection or defining domain-specific similarity function, which further increases its versatility and suitability for a wide range of scenarios.
    
    \item We have developed a dedicated renderer for diff results, which enables users to conveniently visualize and analyze the differences identified by our JSON diff framework.
\end{itemize}

These contributions collectively demonstrate the value of our JSON diff framework as a powerful and flexible tool for JSON diff in various applications and contexts.

The structure of this paper is as follows:

First, we provide a survey of the related works, highlighting the theoretical foundations underpinning the current state-of-the-art and practical implementations found in open-source communities. 

Second, we overview the architecture of our proposed framework, outlining the design principles and key components that contribute to its efficacy and extendability. And we introduce various concepts, assumptions, and notations that will enable us to clearly define the problem and maintain consistency throughout our subsequent discussions. By employing these tools, we can formulate the JSON diff problem and provide the proposed framework.

Following this, we describe the design and implementation of our framework, detailing the relevant algorithms and their applications in real-world scenarios. As we discuss these algorithms, we also demonstrate how they can be adapted to suit the specific requirements of different use cases.

Finally, we conclude the paper with a summary of our findings and a discussion of our research's implications and potential future directions. This section summarizes the key takeaways from our study and provides insights into areas that warrant further exploration, ultimately contributing to the ongoing advancement of the research field.

\section{Related Works}

The study of diff structured data has received widespread attention for years. One notable research paper, \cite{10.1145/235968.233366}, examined the problem of efficiently obtaining the shortest delta operation given a tree structure. The paper focused on two key issues: effectively representing and detecting changes in hierarchical data and utilizing these changes to optimize data synchronization and version control processes. The paper proposed an algorithm that uses a top-down approach to compare the two trees. Starting from the root node, the algorithm systematically compares child nodes until a minimum edit script is found, which can transform one tree into another and provides a valuable foundation for further research in this field.

Besides, \cite{10.1145/3514221.3517850} extensively discussed the algorithms used for entire JSON diffs, providing a detailed analysis of algorithmic complexity and discourse. \cite{10.1145/3514221.3517850} standardized the calculation of "similarity" using the concept of edit distance, and its algorithmic complexity surpassed that of previous frameworks. Compared to this paper, our work focuses more on introducing the implementation of JYCM and demonstrating its high customizability and visualization capabilities.

Furthermore, subsequent researches, such as \cite{xdiff} and \cite{htmldiff}, have examined best practices for working with specific data structures like XML and HTML. In addition, open-source communities have contributed many excellent JSON diff frameworks, such as \cite{github/andreyvit/json-diff} and \cite{github/yudai/gojsondiff}  and \cite{github/zgrossbart/jdd}, which implement various algorithms with different contributions and focuses. For example, \cite{github/andreyvit/json-diff} provides a comprehensive JSON diff framework and includes a deephash library, while \cite{github/yudai/gojsondiff} has made significant progress in fuzzy matching.

\section{Overview and Preliminary}
\subsection{JSON}
JSON, short for JavaScript Object Notation, is a lightweight data-interchange format that is human-readable and machine-readable. It has become a widely adopted standard for data exchange between web applications often used in RESTful APIs, configuration files, and data storage and servers due to its simplicity and compatibility with various programming languages \cite{crockford2006json}. 
A JSON object is an unordered collection of key-value pairs enclosed in curly braces (\{ \}). The keys are strings, and the values can be strings, numbers, booleans, null, objects, or arrays. Objects can be nested within one another, providing a flexible way to model complex data structures.

A JSON array is an ordered collection of values enclosed in square brackets ([ ]). The values within an array can be any valid JSON data type, including objects, arrays, strings, numbers, booleans, or null. Arrays can also be nested within one another to represent multi-dimensional data structures.

JSON supports several primitive data types as shown in Code~\ref{lst:examplejson}:
\begin{itemize}
\item String: A sequence of Unicode characters enclosed in double quotes (" ").
\item Number: A numeric value can be an integer or a floating-point number. JSON does not differentiate between the two.
\item Boolean: Represents true or false values.
\item Null: Represents an empty or non-existent value.
\end{itemize}

\begin{lstlisting}[language=json,xleftmargin=4.5ex,caption={Example JSON from \cite{crockford2006json}},captionpos=b,label={lst:examplejson}]
{
  "Image": {
      "Width":  800,
      "Height": 600,
      "Title":  "View from 15th Floor",
      "Thumbnail": {
          "Height": 125,
          "Width":  "100"
      },
      "IDs": [116, 943, 234, 38793]
    }
}
\end{lstlisting}

\subsection{Design factors}

In designing our framework, we primarily considered several factors:

\begin{itemize}
  \item High coverage: The diff functionality should encompass a wide range of scenarios.
  \item High extensibility and ease of use: The framework must allow users to define scenarios flexibly.
  \item Friendly UI: Human-readable results should also be easy to analyze, even for large JSON files.
\end{itemize}

We divided the diff capabilities according to the components of a JSON, which include Primitive components, Dictionary objects, and Array objects. For each component, we provide various diff strategies tailored to the specific component type.

By incorporating different diff strategies for each JSON component and supporting nested structures, our framework achieves high coverage and extensibility, enabling users to define scenarios flexibly and efficiently.

In addition to the component-specific diff strategies, we adopted a similarity-based design. This approach enables a more nuanced comparison between objects beyond a simple binary distinction of "identical" or "different". As a result, the framework's extensibility and flexibility are  enhanced, providing users with more granular control over the comparison process.

Since our algorithm is recursive, it is essential to define the terminal state, in which no further recursion is required, and the actual algorithm execution can take place.

\subsection{Similarity}

In our design, a single terminal state exists: Given two objects, a similarity score can be calculated between them. The similarity $\Phi$ is defined as follows: a scalar real value ranging from 0 to 1, where a value of 1 indicates complete equality between the objects and 0 indicates complete inequality, respectively, whose formula can be found at \eqref{eq:def:similarity}
\begin{equation} \label{eq:def:similarity}
	\Phi: x, y \rightarrow [0, 1]
\end{equation}
where
$$ x, y \in \{\textbf{STR}, \textbf{NUM}, \textbf{NULL}, \textbf{BOOL}, \textbf{OBJ}, \textbf{ARR}\} $$
for simplicity, we point out the the $\Phi$ should be symmetric, that is 
\begin{equation} \label{eq:similarity:sym}
	\Phi(x,y) = \Phi(y,x)
\end{equation}
and by default, we define 
\begin{equation} \label{eq:similarity:sym}
	\Phi(x,\textbf{NONE}) = 0
\end{equation}
where $\textbf{NONE}$ is for non-existing value.

\subsection{JSON path}
In our approach, we utilize JSON path notation to effectively locate elements within JSON objects. While there is no official standard for JSON path, the basic concepts are widely shared, as outlined in \cite{xpathforjson} and \cite{jsonpointer}, and implemented in \cite{github/json-path/JsonPath} and \cite{github/dchester/jsonpath}. JSON objects exhibit a tree-like structure, allowing each node to be accessed by tracing the path from the root node to the desired target node.

To represent these paths, we adopt a unique symbol ($\rightarrow$) to connect nodes sequentially along the path, ultimately referencing the target element. Furthermore, to offer a more intuitive depiction of a node's position within an array object, we include the array index within square brackets, denoted as $[index]$. This notation enhances the overall readability and comprehension of JSON paths in our framework, providing a clearer understanding of a node's location and its relation to surrounding elements.

We have also extended our JSON path notation to support regular expressions, allowing for more flexible and powerful pattern matching when locating elements within JSON objects. This enables users to find and target nodes based on specific patterns and conditions, improving the versatility and adaptability of JSON paths in our framework.

\begin{lstlisting}[language=json,caption={Example JSON for JSON path},captionpos=b,label={lst:examplejsonforjsonpath}]
{
  "a": 1,
  "b": [
    { "c": 1 },
    { "d": 2 }
  ]
}
\end{lstlisting}

In Table \ref{table:jsonpath} We provide an example of JSON path and the values it receives given Code~\ref{lst:examplejsonforjsonpath}

\begin{table}[htbp]
\caption{Retrieve value by JSON path on Code~\ref{lst:examplejsonforjsonpath}}
\begin{center}
\begin{tabular}{|c|c|}
\hline
\textbf{JSON PATH}& \textbf{VALUE} \\
\hline
$a$ & $1$ \\
\hline
$b \rightarrow [0]$ &  \text{\{ "c": 1 \}} \\
\hline
$b \rightarrow [*]$ &  \text{\{ "c": 1 \} \textbf{and} \{ "d": 2 \}} \\
\hline
$b \rightarrow [0] \rightarrow d$ &  $2$ \\
\hline

\end{tabular}
\label{table:jsonpath}
\end{center}
\end{table}

JSON path empowers users by enabling them to define custom similarity functions for comparing two objects by checking the current differed objects' JSON path as shown in Code~\ref{lst:pycustomsimilarity}. It also gives users an accessible and efficient way to analyze the diff results. By offering this level of control, our framework caters to the specific needs of users, allowing for more precise and meaningful comparisons within the context of their applications.

\subsection{Pairing}
In addition to the primary task of computing the differences between JSON objects, our framework also addresses the challenge of rendering and collecting the diff results. To achieve this, it is crucial to record the optimal similarity pairs, described in JSON path, identified during the execution of the diff algorithm.

In our framework, recording the specific operations is of utmost importance. For example, We must document the transformation process during our diff algorithm, which converts the array $x$ into the array $y$: which elements need to be deleted, added, modified, and preserved. By tracking these operations, we can not only determine the similarity between JSON objects but also provide a clear and concise representation of the changes that have occurred.

Moreover, this approach allows for a more in-depth analysis of the paired JSON objects. For example, by examining the JSON diff, users can gather insights into the primary locations of differences based on JSON path patterns. This information can provide valuable insights to users and guide them in identifying the key areas of change between the JSON objects.

Given two JSON objects $A$ and $B$, we use $\theta$ to denote the set of optimal similarity pairs identified during the execution of the diff algorithm. Each pair in $\theta$ consists of elements (or pointers to those elements) from $A$ and $B$. 

\subsection{Formulation}
With the above definitions and notions, we can now formally describe this JSON diff problem as an optimized problem: give two JSON objects $x$ and $y$, we want to find the paring that maximizes the similarity of these two objects, which can be expressed as in \eqref{eq:def:opt-algorithm}.
\begin{equation} \label{eq:def:opt-algorithm}
    \max_{\theta} \Phi \Big(x,y;\theta\Big)
\end{equation}
And if we donate
$$\theta^* =  \mathop{\arg\max}_{\theta} \Phi \Big(x,y;\theta\Big) \text{ , } \phi^* = \Phi \Big(x,y;\theta^*\Big)$$

Then our diff algorithm $\tilde A$ can be expressed as in \eqref{eq:def:diff-in-math}
\begin{equation} \label{eq:def:diff-in-math}
    \tilde A: x,y \rightarrow \theta^*, \phi^*
\end{equation}

\section{Design and Implementation}

\subsection{Primitive Similarity}
For primitive data types (such as strings, numbers, and boolean values), the default similarity function is relatively straightforward, comparing their equality as shown in Algrithm~\ref{alg:defaultprimitivesimilarity}. However, users can also customize the similarity function by hooking into this functionality, for instance, by utilizing the edit distance to calculate the similarity between two strings. This flexibility allows for more tailored comparisons that cater to the specific needs of the users and their datasets.

\begin{algorithm}[H]
\caption{Default Primitive Similarity}\label{alg:defaultprimitivesimilarity}
\begin{algorithmic}[1]
\Procedure{$\Phi$}{A, B}
   \If{$A == B$}
        \State \Return $1$
   \Else
        \State \Return $0$
   \EndIf
\EndProcedure
\end{algorithmic}
\end{algorithm}

\subsection{Object Similarity}

For the similarity function of JSON objects, the default similarity function computes the average similarity score for each key-pair in objects $A$ and $B$ shown in Algrithm~\ref{alg:defaultobjectsimilarity} where \textbf{keys} is to retrieve all keys of an object (dictionary). This approach takes into consideration the individual similarity scores for each corresponding key-pair, ultimately producing an overall average score that represents the similarity between the two JSON objects under comparison.

\begin{algorithm}[H]
\caption{Default Object Similarity}\label{alg:defaultobjectsimilarity}
\begin{algorithmic}[1]
\Procedure{$\Phi$}{A, B}
    \State Initialize $allKeys \gets \textbf{keys}(A) \cup \textbf{keys}(B)$
    \State Initialize $score \gets 0$

    \For{$key \textbf{ in } allKeys$}
       \If{$A \textbf{ has } key \textbf{ and }  B \textbf{ has } key$}
        \State $score \gets score + \Phi(A[key], B[key]) $
      \ElsIf{$A \textbf{ has } key$}
        \State $score \gets score + \Phi(A[key], \textbf{NONE}) $
       \Else
        \State $score \gets score + \Phi(B[key], \textbf{NONE}) $
       \EndIf
    \EndFor
    \State \Return $socre \text{ / } \textbf{len}(allKeys)$
\EndProcedure
\end{algorithmic}
\end{algorithm}

\subsection{Arrary Similarity}

Array comparison in JSON data is classified into two main categories in our framework: Ordered and Unordered comparisons. Furthermore, each category can be divided into two subcategories: Exact matching and Fuzzy matching. 

The reason why we design this way is as follows.

JSON arrays are generally considered to have an order. However, to accommodate a broader range of scenarios, we allow users to request that the framework treat arrays as unordered "sets" when comparing them. 

When comparing arrays with distinct orderings, such as \textit{(a, b, c)} and \textit{(c, b, a)}, different conclusions may be reached depending on whether the order is considered or not. Therefore, it is essential to treat these cases separately.

The distinction between precise and fuzzy matching is crucial for our framework, as fuzzy matching is fundamentally a pairing problem to find a combination of pairs with the minimum cost. Using the ordered arrays \textit{(a, b, c)} and \textit{(a, z, c)} as an example, it is possible that the cost of matching \textit{b} and \textit{z} is too high, causing the algorithm to incorrectly pair \textit{a} with \textit{z}. This result may not be reasonable in some scenarios, hence the need for fuzzy matching separately.

It is important to note that our work does not introduce any fundamentally new algorithms. Instead, our main contribution lies in combining and adapting existing algorithms to suit common web application JSON-used scenarios. Consequently, we will not provide proof for the fundamental algorithms but will focus on our definitions and context in the framework of these algorithms.

All the fundamental algorithms we use have been summarized together under different matching scenarios Table \ref{table:mafa};

\begin{table}[htbp]
\caption{Matching algorithm for array}
\begin{center}
\begin{tabular}{|c|c|c|}
\hline
 & \textbf{\textit{Exact matching}}& \textbf{\textit{Fuzzy matching}} \\
\hline
\textbf{\textit{Ordered}} & LCS \cite{878178} &  Edit distance \cite{editdistance} \\
\hline
\textbf{\textit{InOrdered}} & Brute force &  Hungarian \cite{Hungarian} \\
\hline

\end{tabular}
\label{table:mafa}
\end{center}
\end{table}

By default, we utilize Algorithm~\ref{alg:defaultarraysimilarityhelper} to compute the similarity between two arrays. This algorithm takes two input arrays, and the pairs obtained through various matching methods. Consequently, the subsequent algorithms discussed in this paper are primarily employed for matching purposes. Once the matching process is completed, we use this function to calculate the real-valued similarity between the arrays.

\begin{algorithm}[H]
\caption{Array Similarity Helper}\label{alg:defaultarraysimilarityhelper}
\begin{algorithmic}[1]
\Procedure{$\Phi_{\textbf{arrayHelper}}$}{A, B, pairs}
    \State Initialize $score \gets 0$
    \State Initialize $n \gets \textbf{len}(A) + \textbf{len}(B)$

    \State \textbf{record}(A,B,pairs)
    
    \For{$pair \textbf{ in } pairs$}
       \State $score \gets score + \Phi(pair[0], pair[1])$
    \EndFor
    \State \Return $socre \text{ / } n$
\EndProcedure
\end{algorithmic}
\end{algorithm}

\subsubsection{Ordered Array similarity under Exact Matching}

For this matching type, we use the Longest Common Subsequence (LCS) \cite{878178} algorithm to find the longest common subsequence between two arrays.

The Longest Common Subsequence (LCS)  algorithm is a dynamic programming method to find the longest subsequence common to two sequences. In the context of matching elements from two ordered array, the LCS algorithm can identify the longest subsequence of elements shared by two arrays, taking into account their order but not necessarily their contiguity. This method is particularly useful for ordered, exact array comparisons where the elements' relative positions matter.

\begin{algorithm}[H]
\caption{Ordered Array Exact Matching}\label{alg:oaem}
\begin{algorithmic}[1]
\Procedure{$\Phi$}{A, B}
    \State Initialize $dp \gets \textsc{LCS}(A,B)$
    \State Initialize $pairs \gets \textsc{BacktrackLCS}(A,B,dp)$
    \State \textbf{return} \textsc{$\Phi_{\textbf{arrayHelper}}$} $(A, B, pairs)$
\EndProcedure
\end{algorithmic}
\end{algorithm}

The whole algorithm we used here is described in Algorithm~\ref{alg:oaem}, which is composed of two parts: first we apply the LCS, described in Algorithm~\ref{alg:lcs}, then we use another procedure, described in 
 Algorithm~\ref{alg:lcsbacktrack}, to backtrack what are exactly the common elements.

\begin{algorithm}[H]
\caption{Longest Common Subsequence}\label{alg:lcs}
\begin{algorithmic}[1]
\Procedure{LCS}{A, B}
    \State Initialize $n \gets \text{length}(A)$
    \State Initialize $m \gets \text{length}(B)$
    \State Initialize $dp[0 \dots n, 0 \dots m]$ with all zeros
    \For{$i = 1 \textbf{ to } n$}
        \For{$j = 1 \textbf{ to } m$}
            \If{$1 == \Phi(A[i], B[j]) $}
                \State $dp[i, j] \gets dp[i-1, j-1] + 1$
            \Else
                \State $dp[i, j] \gets \max(dp[i-1, j], dp[i, j-1])$
            \EndIf
        \EndFor
    \EndFor
    \State \textbf{return} $dp$
\EndProcedure
\end{algorithmic}
\end{algorithm}

\begin{figure}[H]
\centering
\includegraphics[width=0.9\textwidth]{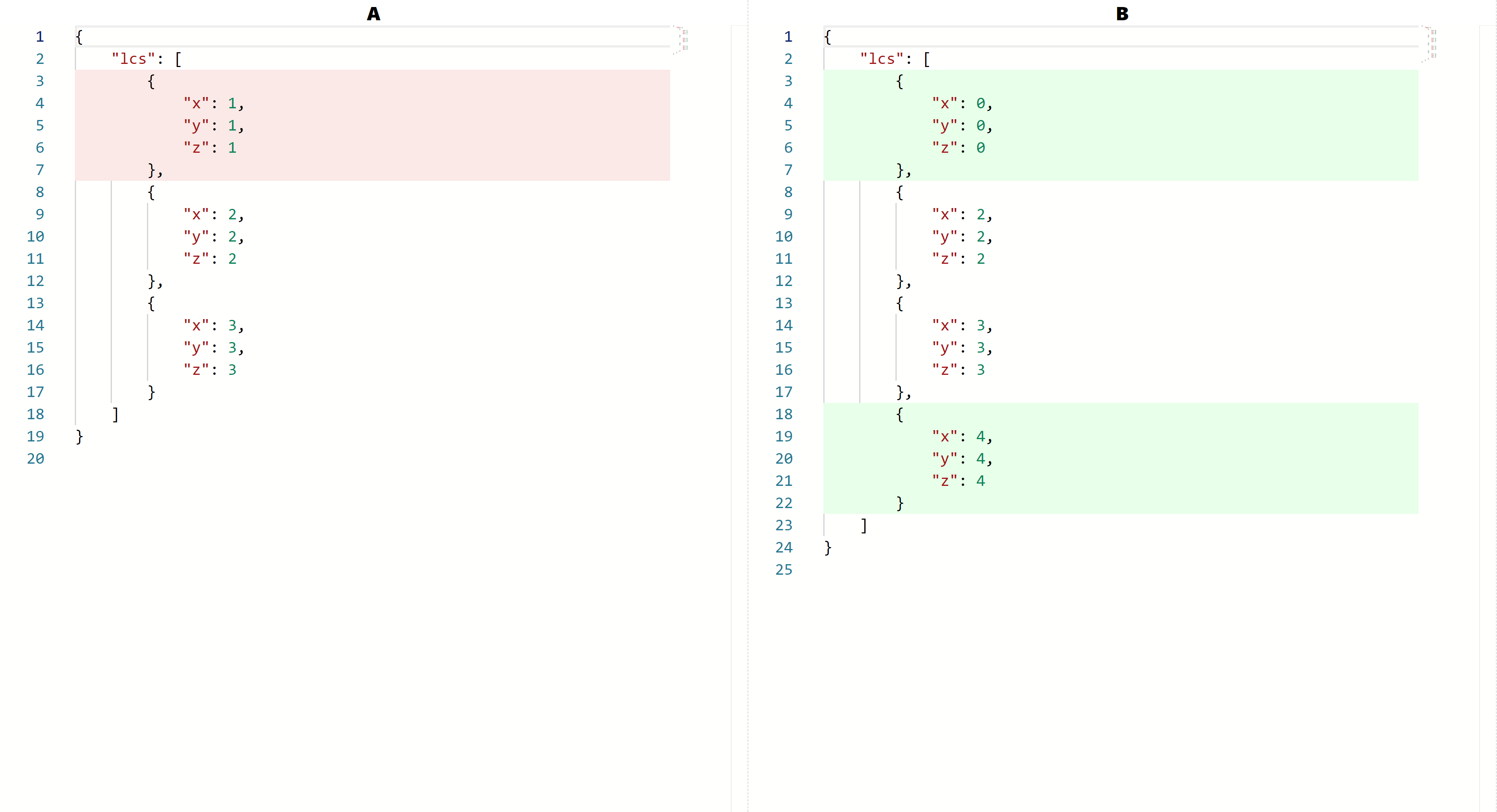}
\caption{Ordered Array Exact Matching}
\label{fig:jycmoaem}
\end{figure}

As shown in Fig.~\ref{fig:jycmoaem}, this type of matching is useful when the order of the elements is crucial, and only identical matches are considered valid. One such application is in the field when swapping the order of operations or introducing different events could result in unexpected behavior or even errors. An example of this would be comparing the outputs of two APIs that provide lists of chronological events, such as user activity logs or transaction histories, where the order of events is essential and the exact details of each event need to match.

\begin{algorithm}[H]
\caption{LCS backtrack}\label{alg:lcsbacktrack}
\begin{algorithmic}[1]
\Procedure{BacktrackLCS}{A, B, dp}
    \State Initialize $pairs$ as an empty list
    \State Initialize $i \gets \text{length}(A)$
    \State Initialize $j \gets \text{length}(B)$
    \While{$i > 0$ and $j > 0$}
        \If{$X[i-1] == Y[j-1]$}
            \State Prepend $[i-1,j-1]$ to $pairs$
            \State $i \leftarrow i - 1$, $j \leftarrow j - 1$
        \ElsIf{$dp[i-1][j] > dp[i][j-1]$}
            \State $i \leftarrow i - 1$
        \Else
            \State $j \leftarrow j - 1$
        \EndIf
    \EndWhile
    \State \textbf{return} $pairs$
\EndProcedure

\end{algorithmic}
\end{algorithm}

\subsubsection{Ordered Array similarity under Fuzzy Matching}

In ordered fuzzy matching, we use a variation of the Edit Distance algorithm \cite{editdistance} to find the minimum cost matching between two arrays. In this case, the default cost of "editing" two elements is the negative of their similarity; that is, the more they are similar, the less they need to be edited.

The Edit Distance algorithm, also known as the Levenshtein distance, is a dynamic programming technique used to determine the minimum number of edit operations required to transform one sequence into another. 

\begin{algorithm}[H]
\caption{Ordered Array Fuzzy Matching}\label{alg:oafm}
\begin{algorithmic}[1]
\Procedure{$\Phi$}{A, B}
    \State Initialize $dp \gets \textsc{EditDistance}(A,B)$
    \State Initialize $pairs \gets \textsc{BacktrackEditDistance}(A,B,dp)$
    \State \textbf{return} \textsc{$\Phi_{\textbf{arrayHelper}}$} $(A, B, pairs)$
\EndProcedure
\end{algorithmic}
\end{algorithm}

The whole algorithm we used here is described in Algorithm~\ref{alg:oafm} which is composed of two parts: first, we apply the LCS, described in Algorithm~\ref{alg:lcs}, then we use another procedure, described in 
 Algorithm~\ref{alg:lcsbacktrack} where $\textbf{zeros}$ is a helper function to create a matrix filled with zeros, to backtrack what are exactly the common elements.

In the context of array comparison, the Edit Distance algorithm can quantify the similarity between two arrays by calculating the minimum number of element insertions, deletions, and substitutions needed to make the arrays identical. 

This method is particularly useful for ordered, approximate array comparisons where the elements' relative positions matter.

\begin{algorithm}[H]
\caption{Edit Distance}\label{alg:editdistance}
\begin{algorithmic}[1]

\Procedure{EditDistance}{A, B}

\State Initialize $m \gets 1 + \textbf{len}(A)$
\State Initialize $n \gets 1 + \textbf{len}(B)$

\State Initialize $dp \gets \textbf{zeros}(m,n)$

\For{$x \in \{m - 2, \dots, 0\}$}
    \For{$y \in \{n - 2, \dots, 0\}$}
        \State $dp[x][y]$ $\gets$ $\max$( \par
        \hskip\algorithmicindent \hskip\algorithmicindent $dp[x + 1][y]$, \par
        \hskip\algorithmicindent \hskip\algorithmicindent $dp[x][y + 1]$, \par
        \hskip\algorithmicindent \hskip\algorithmicindent $\Phi(A[x], B[y]) + dp[x + 1][y + 1])$\par
        \hskip\algorithmicindent)

    \EndFor
\EndFor
\State \textbf{return} $dp$
\EndProcedure

\end{algorithmic}
\end{algorithm}

\begin{algorithm}[H]
\caption{Edit Distance Backtrack}\label{alg:editdistancebacktrack}
\begin{algorithmic}[1]
\Procedure{EditDistanceBacktrack}{A, B}

\State Initialize $m \gets 1 + \textbf{len}(A)$
\State Initialize $n \gets 1 + \textbf{len}(B)$
\State Initialize $pairs$ as an empty list
\State Initialize $i \gets 0$
\State Initialize $j \gets 0$

\While{$i + j < m + n - 2$}
    \State $curr \leftarrow dp[i][j]$
    \State $pxs \leftarrow 0$ if $i + 1 \geq m$ else $dp[i + 1][k]$
    \State $pys \leftarrow 0$ if $j + 1 \geq n$ else $dp[i][j + 1]$

    \If{$curr == pxs$ and $i + 1 < m$} \label{lst:editdistancebacktrack:byremoving}
        \State $i \leftarrow i + 1$
        \State \textbf{continue}
    \EndIf

    \If{$curr == pys$ and $j + 1 < n$} \label{lst:editdistancebacktrack:byadding}
        \State $j \leftarrow j + 1$
        \State \textbf{continue}
    \EndIf

    \State Append $[i,j]$ to $pairs$

    \State $i \leftarrow i + 1$
    \State $j \leftarrow j + 1$
\EndWhile

\State \textbf{return} $pairs$
\EndProcedure

\end{algorithmic}
\end{algorithm}

Just for clarification, to re-construct from $A[i]$ to $B[j]$,  line~\ref{lst:editdistancebacktrack:byremoving} is for removing $A[i]$
 and line~\ref{lst:editdistancebacktrack:byadding} is for adding $B[j]$

\begin{figure}[H]
\centering
\includegraphics[width=0.9\textwidth]{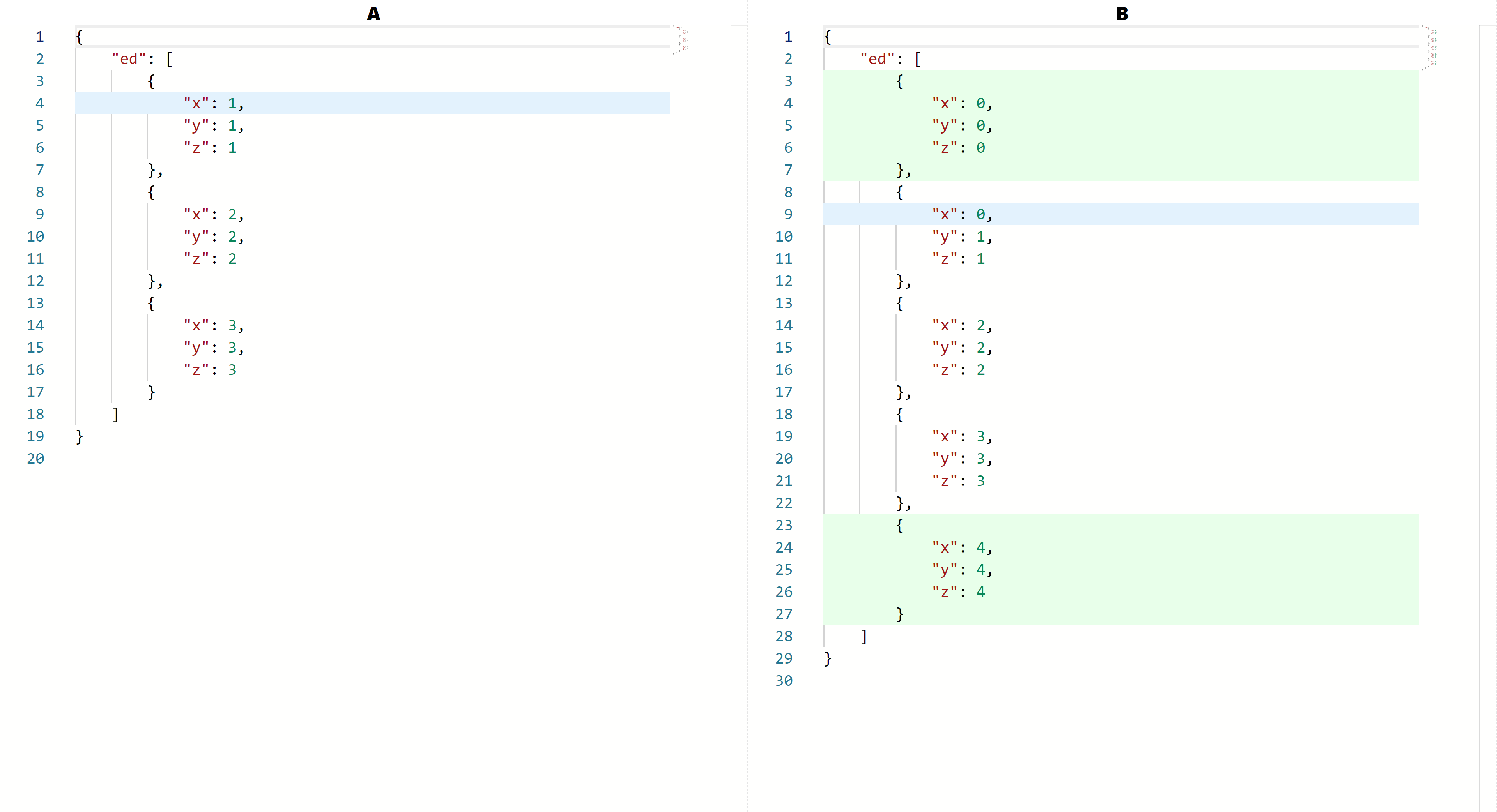}
\caption{Ordered Array Fuzzy Matching}
\label{fig:jycmoafm}
\end{figure}

As shown in Fig.~\ref{fig:jycmoafm}, this type of matching is useful when the order of the elements is essential, but some degree of flexibility is allowed in terms of matching the elements themselves. One such application is in the field of natural language processing, where the order of words or phrases is significant, but synonyms or paraphrasing can still convey the same meaning.
An example of this would be comparing the outputs of two APIs that provide ranked lists of search results, such as product listings or top news articles, where the order of the results is essential, but the exact details of each result might vary slightly.

\subsubsection{Unordered Array similarity under Exact Matching}

We use a brute-force approach to find matching pairs between two arrays in unordered exact matching (i.e., two sets). Usually, we should use hash to deal with such tasks effectively. However considering the difficulties of calculating hash value in the context of flexibility and user-defined similarity, taking the IOU example in the previous section it is impossible to use a reasonable hash function to hash those two high-IOU-yet-different coordinates into the same value, we use a two-depth nested for-loop, 

The procedure is described in Algorithm~\ref{alg:uaem} where the brute force approach is described in Algorithm~\ref{alg:bruteforce}

\begin{algorithm}[H]
\caption{Unordered Array Exact Matching}\label{alg:uaem}
\begin{algorithmic}[1]
\Procedure{$\Phi$}{A, B}
    \State Initialize $pairs \gets \textsc{BruteForceMatching}(A,B)$
    \State \textbf{return} \textsc{$\Phi_{\textbf{arrayHelper}}$} $(A, B, pairs)$
\EndProcedure
\end{algorithmic}
\end{algorithm}

\begin{algorithm}[H]
\caption{BruteForceMatching}\label{alg:bruteforce}
\begin{algorithmic}[1]
\Procedure{BruteForceMatching}{A, B}

\State Initialize $m \gets \textbf{len}(A)$
\State Initialize $n \gets \textbf{len}(B)$
\State Initialize $i \gets 0$
\State Initialize $j \gets 0$

\For{$i = 1 \textbf{ to } m$}
    \For{$j = 1 \textbf{ to } n$}
        \If{$1 == \Phi(A[i], B[j]) $}
            \State Append $[i,j]$ to $pairs$
        \EndIf
    \EndFor
\EndFor

\State \textbf{return} $pairs$
\EndProcedure

\end{algorithmic}
\end{algorithm}

\begin{figure}[H]
\centering
\includegraphics[width=0.9\textwidth]{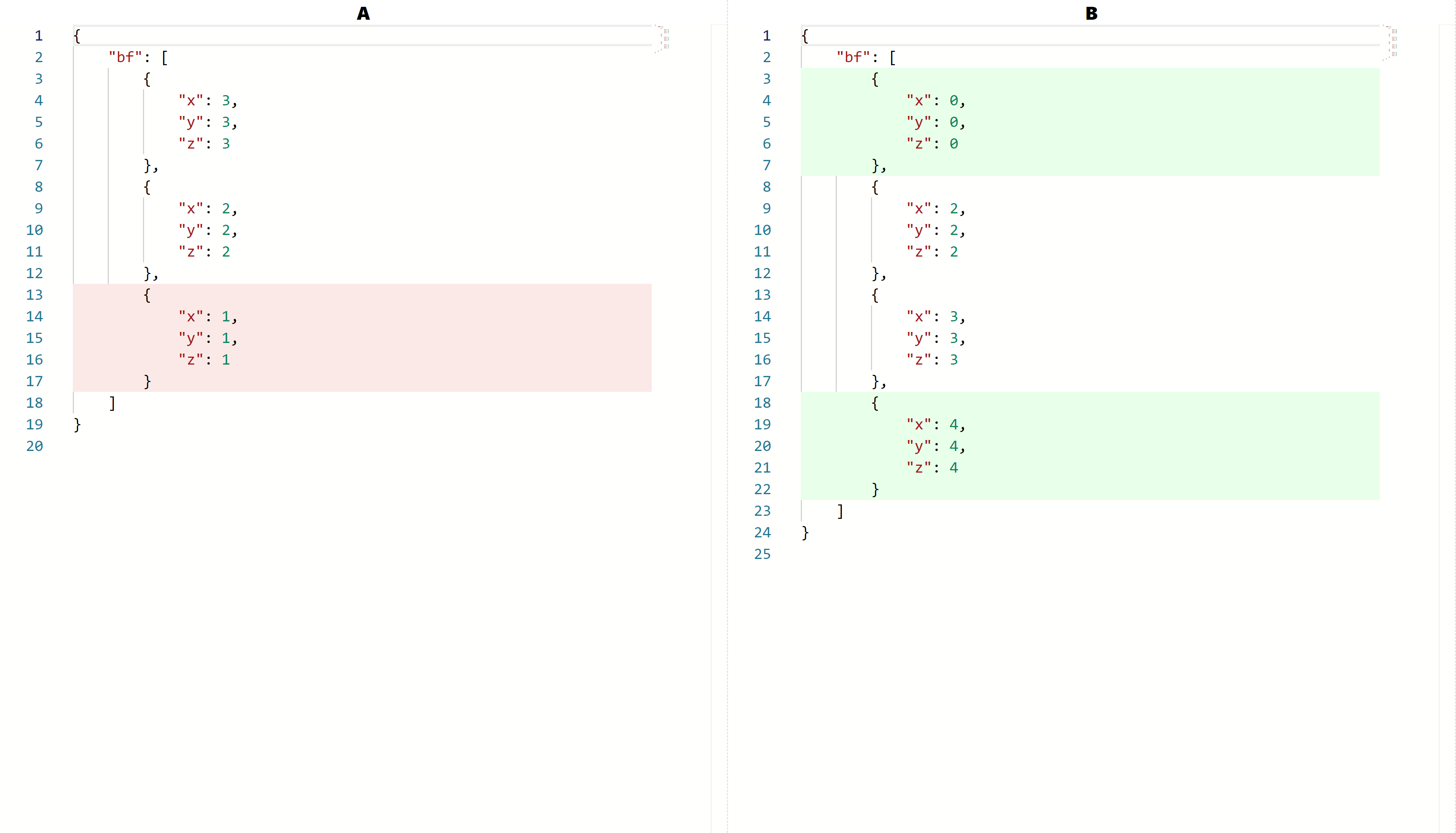}
\caption{Unordered Array Exact Matching}
\label{fig:jycmuaem}
\end{figure}

As shown in Fig.~\ref{fig:jycmuaem}, an ideal scenario for this matching algorithm would be when the order of the elements is not essential, but only identical matches are considered valid. One such application is in the field of inventory management, an example of this would be comparing the outputs of two inventory management APIs that return lists of items in stock, where the order of items is not important, but the exact items and their properties need to match.

\subsubsection{Unordered Array similarity under Fuzzy Matching}

Before delving into this scenario, we first formulate this matching problem as follows: Given two sets of elements, where the similarity between any two elements can be calculated using \eqref{eq:def:similarity}, we aim to find a pairing method that maximizes the \textbf{total similarity} of the pairings.

To achieve this, we must define the \textbf{total similarity}. By default, we employ Algorithm~\ref{alg:defaultarraysimilarityhelper} to compute it.

This problem formulation aligns well with the Hungarian algorithm \cite{Hungarian}, also known as the Kuhn-Munkres algorithm. This efficient method solves the assignment problem, which involves assigning tasks to agents in a manner that minimizes the total cost of the assignments.

Due to space constraints, we will not provide a detailed explanation of this algorithm. However, we shall define its input and output. Given an $m \times n$ cost matrix $costMatrix = [costMatrix_{ij}]$, where $costMatrix_{ij}$ represents the cost of assigning the $i$-th worker to the $j$-th job, the goal is to find a permutation $\sigma$ that minimizes the total cost $\sum_{i=1}^{n} c_{i, \sigma(i)}$. This is where the \textbf{hungarian} function comes into play, as described in \eqref{eq:def:hungarian}, where $\sigma(i)$ can be deduced from the $pairs$ variable:

\begin{equation} \label{eq:def:hungarian}
\textbf{hungarian}: costMatrix \rightarrow pairs
\end{equation}

The algorithm operates by constructing a cost matrix representing the dissimilarity between each pair of elements in the two sets. It then iteratively modifies the cost matrix by subtracting the smallest element in each row and column until a complete set of assignments can be made with zero total cost. The optimal matching is obtained from the modified cost matrix by identifying the unique assignments corresponding to zero-cost pairs.

We can now describe our matching algorithm under this scenario in Algorithm~\ref{alg:uafm}

\begin{algorithm}[H]
\caption{Unordered Array Fuzzy Matching}\label{alg:uafm}
\begin{algorithmic}[1]
\Procedure{$\Phi$}{A, B}
    \State Initialize $costMatrix \gets -1 \times$ $sm ^ *$
    \State \textbf{return} \textsc{$\Phi_{\textbf{arrayHelper}}$} $(A, B, \textbf{hungarian}(costMatrix))$
\EndProcedure
\end{algorithmic}
\end{algorithm}

$^*sm$ is calculated as below:

\[
sm =
\begin{blockarray}{cccc}
\begin{block}{[cccc]}
  \Phi(A_1,B_1) & \Phi(A_1,B_2) & \cdots & \Phi(A_1,B_n) \\
  \Phi(A_2,B_1) & \Phi(A_2,B_2) & \cdots & \Phi(A_2,B_n) \\
  \vdots & \vdots & \vdots & \vdots \\
  \Phi(A_n,B_1) & \Phi(A_m,B_2) & \cdots & \Phi(A_m,B_n) \\
\end{block}
\end{blockarray}
 \]

\begin{figure}[H]
\centering
\includegraphics[width=0.9\textwidth]{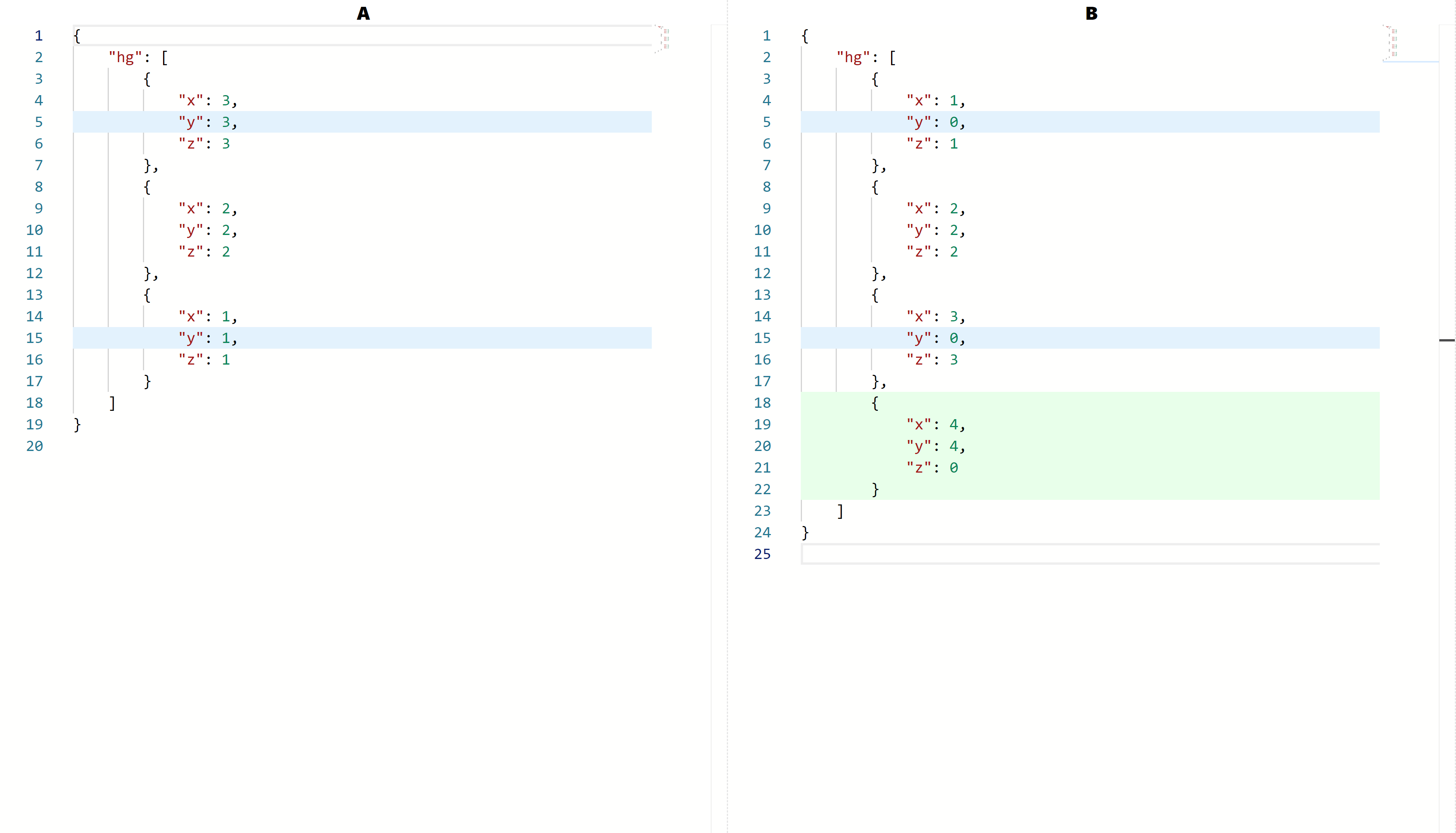}
\caption{Unordered Array Fuzzy Matching}
\label{fig:jycmuafm}
\end{figure}

As shown in Fig.~\ref{fig:jycmuafm}, an ideal scenario would be when the order of the elements is not essential, and some degree of flexibility is allowed in terms of matching the elements themselves. An example of this would be comparing the output of two search engine APIs that return similar but not identical results.

\subsection{Renderer}

We have utilized React\cite{github/facebook/react}, an open-source front-end library, to implement our renderer. Leveraging the context feature of React, we have made it easy for users to override our rendering logic, such as the coloring scheme and the presentation of specific diff information. For example, users can customize the rendering of the edit distance between two strings according to their preferences. The code for our renderer can be found in the repository at \href{react-jycm-viewer}{react-jycm-viewer}, which you can access on  \href{(https://github.com/eggachecat/react-jycm-viewer}
Furthermore, our renderer supports the display of large JSON objects and seamless navigation between pairings. This is made possible by the integration of monaco-editor\cite{github/microsoft/monaco-editor} project, which enables efficient browsing and searching within large JSON files. Thanks to this feature, our JYCM renderer can handle substantial JSON objects and diff results.

\section{Conclusion}
This paper presents a comprehensive framework for comparing and analyzing JSON objects by identifying their differences. Our approach emphasizes the computation of optimal similarity between JSON objects and the rendering of diff results in a user-friendly manner, accommodating various scenarios and empowering users to define custom similarity functions that fit within the framework easily. As discussed, no universal rule defines "what has been changed" without considering realistic scenarios. We have employed JSON path notation to locate and represent elements within JSON objects and have introduced regular expression support for more flexible path matching. Moreover, our renderer, built using the React library, enables users to customize rendering logic, such as color schemes and diff presentation styles.

We have demonstrated the effectiveness of our framework in various practical scenarios by applying different algorithms, such as the Hungarian algorithm for ordered exact matching and the default array similarity algorithm for unordered matching. Our framework also supports the implementation of user-defined similarity functions for more specific use cases.

\printbibliography

@article{10.1145/235968.233366,
author = {Chawathe, Sudarshan S. and Rajaraman, Anand and Garcia-Molina, Hector and Widom, Jennifer},
title = {Change Detection in Hierarchically Structured Information},
year = {1996},
issue_date = {June 1996},
publisher = {Association for Computing Machinery},
address = {New York, NY, USA},
volume = {25},
number = {2},
issn = {0163-5808},
url = {https://doi.org/10.1145/235968.233366},
doi = {10.1145/235968.233366},
journal = {SIGMOD Rec.},
month = {jun},
pages = {493–504},
numpages = {12}
}

@misc{github/andreyvit/json-diff,
title = {json-diff},
publisher = {GitHub},
journal = {GitHub repository},
howpublished = {\url{https://github.com/andreyvit/json-diff}}
}

@MISC{github/yudai/gojsondiff,
title = {gojsondiff},
publisher = {GitHub},
journal = {GitHub repository},
howpublished = {\url{https://github.com/yudai/gojsondiff}},
}

@MISC{github/facebook/react,
title = {React},
publisher = {GitHub},
journal = {GitHub repository},
howpublished = {\url{https://github.com/facebook/react}},
}

@MISC{github/microsoft/monaco-editor,
title = {monaco-editor},
publisher = {GitHub},
journal = {GitHub repository},
howpublished = {\url{https://github.com/microsoft/monaco-editor}},
}

@MISC{github/zgrossbart/jdd,
title = {jdd},
publisher = {GitHub},
journal = {GitHub repository},
howpublished = {\url{https://github.com/zgrossbart/jdd}},
}

@article{Hungarian,
author = {Kuhn, H. W.},
title = {The Hungarian method for the assignment problem},
journal = {Naval Research Logistics Quarterly},
volume = {2},
number = {1-2},
pages = {83-97},
doi = {https://doi.org/10.1002/nav.3800020109},
url = {https://onlinelibrary.wiley.com/doi/abs/10.1002/nav.3800020109},
eprint = {https://onlinelibrary.wiley.com/doi/pdf/10.1002/nav.3800020109},
year = {1955}
}

@INPROCEEDINGS{878178,
  author={Bergroth, L. and Hakonen, H. and Raita, T.},
  booktitle={Proceedings Seventh International Symposium on String Processing and Information Retrieval. SPIRE 2000}, 
  title={A survey of longest common subsequence algorithms}, 
  year={2000},
  volume={},
  number={},
  pages={39-48},
  doi={10.1109/SPIRE.2000.878178},
  ISSN={},
  month={Sep.},}

@article{editdistance,
author = {Navarro, Gonzalo},
title = {A Guided Tour to Approximate String Matching},
year = {2001},
issue_date = {March 2001},
publisher = {Association for Computing Machinery},
address = {New York, NY, USA},
volume = {33},
number = {1},
issn = {0360-0300},
url = {https://doi.org/10.1145/375360.375365},
doi = {10.1145/375360.375365},
journal = {ACM Comput. Surv.},
month = {mar},
pages = {31–88},
numpages = {58}
}

@inproceedings{pezoa2016foundations,
  title={Foundations of JSON schema},
  author={Pezoa, Felipe and Reutter, Juan L and Suarez, Fernando and Ugarte, Mart{\'\i}n and Vrgo{\v{c}}, Domagoj},
  booktitle={Proceedings of the 25th International Conference on World Wide Web},
  pages={263--273},
  year={2016},
  organization={International World Wide Web Conferences Steering Committee}
}

@INPROCEEDINGS{xdiff,
  author={Wang, Y. and DeWitt, D.J. and Cai, J.-Y.},
  booktitle={Proceedings 19th International Conference on Data Engineering (Cat. No.03CH37405)}, 
  title={X-Diff: an effective change detection algorithm for XML documents}, 
  year={2003},
  volume={},
  number={},
  pages={519-530},
  doi={10.1109/ICDE.2003.1260818}}

@article{htmldiff,
  title={HtmlDiff: a differencing tool for HTML documents},
  author={Berk, Elliot},
  journal={Student Project, Princeton University},
  year={1996}
}

@article{xpathforjson,
  title={JSONPath - XPath for JSON},
  author={Gössner, Stefan},
  url = {https://goessner.net/articles/JsonPath/},
  year={2007}
}

@article{jsonpointer,
  title={JavaScript Object Notation (JSON) Pointer,},
  author={P. Bryan, Ed and K. Zyp and M. Nottingham, Ed},
  url = {https://www.rfc-editor.org/info/rfc6901},
  year={2013}
}

@MISC{github/json-path/JsonPath,
title = {JsonPath},
publisher = {GitHub},
journal = {GitHub repository},
howpublished = {\url{https://github.com/json-path/JsonPath}},
}

@MISC{github/dchester/jsonpath,
title = {jsonpath},
publisher = {GitHub},
journal = {GitHub repository},
howpublished = {\url{https://github.com/dchester/jsonpath}},
}

@misc{crockford2006json,
  title={The application/json Media Type for JavaScript Object Notation (JSON)},
  author={Crockford, Douglas},
  year={2006},
  howpublished={RFC 4627},
  url={https://tools.ietf.org/html/rfc4627}
}

@inproceedings{10.1145/3514221.3517850,
author = {H\"{u}tter, Thomas and Augsten, Nikolaus and Kirsch, Christoph M. and Carey, Michael J. and Li, Chen},
title = {JEDI: These Aren't the JSON Documents You're Looking For...},
year = {2022},
isbn = {9781450392495},
publisher = {Association for Computing Machinery},
address = {New York, NY, USA},
url = {https://doi.org/10.1145/3514221.3517850},
doi = {10.1145/3514221.3517850},
booktitle = {Proceedings of the 2022 International Conference on Management of Data},
pages = {1584–1597},
numpages = {14},
location = {Philadelphia, PA, USA},
series = {SIGMOD '22}
}
\onecolumn
\appendix
\subsection{JYCM result}
\label{apdx:jycmresult}

\subsection{JYCM define custom similarity}
\label{apdx:customsimilarity}

\begin{minipage}{\linewidth}

\begin{lstlisting}[language=Python,numbers=left,caption={Example code to define similarity function within JYCM},captionpos=b,label={lst:pycustomsimilarity}]
import math
from jycm.operator import  BaseOperator

class L2DistanceOperator(BaseOperator):
    __operator_name__ = "operator:l2distance"
    __event__ = "operator:l2distance"

    def __init__(self, path_regex, distance_threshold):
        super().__init__(path_regex=path_regex)
        self.distance_threshold = distance_threshold

    def diff(self, level: 'TreeLevel', instance, drill: bool) -> Tuple[bool, float]:
        distance = math.sqrt(
            (level.left["x"] - level.right["x"]) ** 2 + (level.left["y"] - level.right["y"]) ** 2
        )
        info = {
            "distance": distance,
            "distance_threshold": self.distance_threshold,
            "pass": distance < self.distance_threshold
        }

        if not drill:
            instance.report(self.__event__, level, info)
        return True, 1 if info["pass"] else 0
\end{lstlisting}
\end{minipage}






\end{document}